\documentclass[aps,prl,reprint,twocolumn,groupedaddress,amsmath,amssymb,amsfonts,floats,floatfix,noraggedbottom,nobalancelastpage,dvips,showpacs,10pt]{revtex4}

\usepackage{epsfig} 
\usepackage{psfrag}
\usepackage{braket}
\usepackage{color}
\usepackage{textcomp}
\usepackage{graphicx}
\usepackage{latexsym}
\usepackage{hyperref}
\hypersetup{dvips,
  pdfauthor={Christopher N. Varney, Kai Sun, Victor Galitski, and
    Marcos Rigol},
  pdftitle={Kaleidoscope of exotic quantum phases in a frustrated XY model},
  colorlinks=true,
  linkcolor=blue,
  citecolor=blue,
  pdfpagemode=UseNone
}

\usepackage{breakurl}

\newcommand{\bbraket}[1]{\braket{\hspace{-2pt}\braket{#1}\hspace{-2pt}}}

\begin{document}

\title{Kaleidoscope of exotic quantum phases in a frustrated $XY$ model}

\author{Christopher N. Varney,$^{1,2}$ Kai Sun,$^{1,3}$ Victor
  Galitski,$^{1,3}$ and Marcos Rigol$^2$} 

\affiliation{$^1$Joint Quantum Institute and Department of Physics,
  University of Maryland, College Park, Maryland 20742, USA}
\affiliation{$^2$Department of Physics, Georgetown University,
  Washington, D.C. 20057, USA} 
\affiliation{$^3$Condensed Matter Theory Center, Department of Physics,
  University of Maryland, College Park, Maryland 20742, USA}

\begin{abstract}
  The existence of quantum spin liquids was first conjectured by
  Pomeranchuk some 70 years ago, who argued that frustration in simple
  antiferromagnetic theories could result in a Fermi-liquid-like state
  for spinon excitations. Here we show that a simple quantum spin
  model on a honeycomb lattice hosts the long sought for Bose metal
  with a clearly identifiable Bose surface.  The complete phase
  diagram of the model is determined via exact diagonalization and is
  shown to include four distinct phases separated by three quantum
  phase transitions.
\end{abstract}

\pacs{
  75.10.Kt, 
  67.85.Jk, 
  21.60.Fw, 
  75.10.Jm  
}


\maketitle


We learn early in our education that as matter is cooled down to low
temperatures it normally experiences transitions into ordered states
of various kinds--crystalline solid structures, ordered magnetic
phases, superfluid and superconducting states, etc. It is also common
knowledge that upon heating the matter, the ordered phases melt into
the familiar gaseous or liquid classical states that we encounter
routinely in our everyday lives. A more specialized but equally
well-established result is that no order that breaks a continuous symmetry
can survive in one-dimensional systems~\cite{cazalilla2011},
because quantum zero-point motion acts there similarly to thermal
effects and ``quantum-melts'' ordered phases even at zero temperature.

It has been a long-standing and important question in physics whether
quantum fluctuations in higher-dimensional quantum spin or boson
systems can have the same debilitating effect, giving way to quantum
liquids~\cite{pomeranchuk1941}. The interest in such a hypothetical
spin liquid, also known as a Bose or spin metal~\cite{note1}, has
experienced multiple revivals with the most prominent one associated
with the discovery of high-temperature
superconductivity~\cite{anderson1987, lee2006}. However, despite the
decades of intensive search, no convincing examples of a gapless spin
liquid have been found in any realistic two-dimensional quantum model.

In models that are fermionizable via the Jordan-Wigner
transformation~\cite{jordan1928,galitski2010}, the existence of spin
liquids has now been firmly
established~\cite{paramekanti2002,galitski2010}, but the physics there
mimics somewhat the one-dimensional result~\cite{emery2000,
  vishwanath2001,mukhopadhyay2001}. What remains of crucial importance
is whether a truly higher-dimensional spin system may host a quantum
liquid. Among the influential recent results here are the stability
argument by Hermele {\em et al}.~\cite{hermele2004}, who showed that
there is no fundamental obstacle to the existence of quantum spin
liquids, and a complete classification of quantum orders by
Wen~\cite{wen2002}, who demonstrated that an amazing variety of
hypothetical gapless spin liquids can all be divided into several
distinct classes, which include stable phases with low-lying fermionic
spinon excitations that resemble a Fermi-liquid state. Also, the work
of Motrunich, Fisher, and Sheng~\cite{motrunich2007,sheng2009}
provides strong arguments in favor of the existence of such putative
two-dimensional Bose metals and suggests that the strong singularity
in the spin structure factor at a Bose surface is one of the hallmark
phenomena of this exotic state.

The main idea is that, despite the fact that the underlying particles
are bosons, the collective behaviors in these strongly correlated
Bose-metal states show a strong analogy to a Fermi liquid formed by
fermionic particles. In a Fermi liquid, the fermion statistics dictate
the formation of Fermi surfaces, which possess singular behavior.  In
a Bose metal, despite the absence of Pauli's principle, similar
singularities also arise and define a surface in momentum space, known
as a Bose surface~\cite{motrunich2007,sheng2009}. The existence of a
Bose surface is the key property and most striking experimental
feature of a Bose metal. However, unlike a Fermi liquid, where the
Luttinger theorem requires that the Fermi wave vector depends on the
density of fermions, the Bose wave vector in a Bose metal depends on
the control parameters and can vary continuously even at fixed
particle density.

Here we provide strong evidence that a model as simple as the
$XY$-spin model on a honeycomb lattice with nearest-neighbor (NN) and
next-to-nearest-neighbor (NNN) interactions hosts, among other phases,
a Bose metal with a clearly-identifiable Bose surface. Although, we
came across this finding serendipitously, we would like to provide
qualitative arguments that could potentially guide searches for other
such interesting spin models. Note that the description of a spin
Fermi-liquid-like state is necessarily a gauge
theory~\cite{ioffe1989,lee2006,motrunich2005}, which is very similar
to that of the Halperin-Lee-Read~\cite{halperin1993} quantized Hall
(QH) state. In the latter gapless phase, the interacting electron
system in a large classical external field gives rise to composite
fermions in zero classical field but coupled to a fluctuating quantum
field - the Chern-Simons field that implements flux attachment. The
natural question here, considered before, e.g., in
Refs.~\onlinecite{seidel2005} and \onlinecite{burkov2010}, is whether
a fractional QH state of this or any other type is possible in a
sensible lattice model.

The above remarks are relevant to our work because our simple
Hamiltonian, see Eq.~(\ref{eq:Ham_boson}) below, can be viewed as a
natural ``trial model'' for such a possible fractional lattice QH
state per the following construction. Take the Haldane
model~\cite{haldane1988} of noninteracting electrons on a honeycomb
lattice with simple NN hoppings and complex NNN hoppings,
$|J_2|e^{i\phi}$. If $\phi$ is nonzero it realizes a topological
insulator or lattice ``integer'' QH state. Replace the fermions with
hard-core bosons at half-filling~\cite{varney2010} and it becomes a
promising strongly interacting model. Notably, the most frustrated
limit corresponds to $\phi = \pi$, which maps at half-filling to the
following Hamiltonian
\begin{align}
  H &= J_1 \sum_{\braket{ij}} ( b_i^\dag b_j^{\phantom\dag} + \text{H.c.} ) +
  J_2 \sum_{\bbraket{ij}} ( b_i^\dag b_j^{\phantom\dag} + \text{H.c.} ),
\label{eq:Ham_boson}
\end{align}
where $b_i^\dag$ $(b_i^{\phantom\dag})$ is an operator that creates
(annihilates) a hard-core boson on site $i$. Here, we require the sign
of $J_2$ to be positive (that is, $\phi = \pi$), while the sign of
$J_1$ is in fact irrelevant because of the particle-hole symmetry of
the honeycomb lattice ($b_i\rightarrow -b_i$ for one of the two
sublattices).  In what follows, $J_1= 1$ sets our unit of energy. Note
that Hamiltonian~(\ref{eq:Ham_boson}) maps to a frustrated
anti-ferromagnetic-$XY$ model ($b_i^\dag \to S_i^+$ and
$b_i^{\phantom\dag} \to S_i^-$),
\begin{align}
  H &= J_1 \sum_{\braket{ij}} ( S_i^+ S_j^- + \text{H.c.} ) + J_2
  \sum_{\bbraket{ij}} ( S_i^+ S_j^- + \text{H.c.} ).
  \label{eq:Ham_xy}
\end{align}

\begin{figure}[b]
  \centering
  \includegraphics*[height=\columnwidth,angle=-90,viewport=47 10 458 700,draft=false]{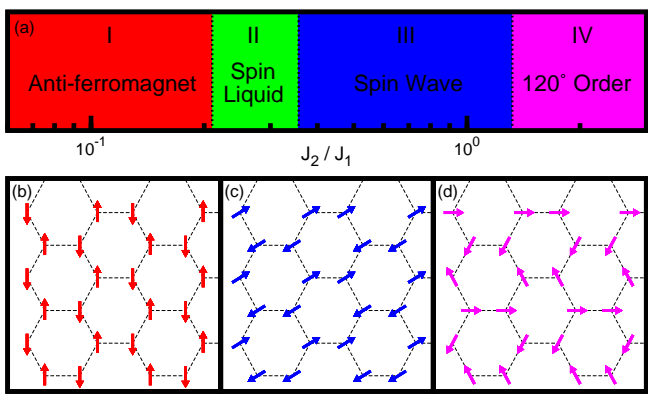}
  \caption{
    (Color online) (a) Phase diagram of the model in
    Eq.~\eqref{eq:Ham_boson} as a function of $J_2 / J_1$, (b)
    antiferromagnetic ordering (phase I), (c) spin wave ordering with
    wavevector ${\bf k } = M$ (phase III), (d) collinear spin wave
    ordering with wavevector ${\bf k} = K$ (phase IV). The phase
    boundaries are $(J_2 / J_1)_{\text{I}\to\text{II}} = 0.210 \pm
    0.008$, $(J_2 / J_1)_{\text{II}\to\text{III}} = 0.356 \pm 0.009$,
    and $(J_2 / J_1)_{\text{III}\to\text{IV}} = 1.320 \pm 0.020$.
    \label{fig:pd}
  }
\end{figure}

The properties of this Hamiltonian are governed by the dimensionless
control parameter $J_2 / J_1$. The limits of this model are well
understood. For $J_2 / J_1 = 0$, the ground state of this Hamiltonian
is an antiferromagnet [Fig.~\ref{fig:pd}(b)]. When $J_2 / J_1 \to
\infty$, however, the ground state is a spin wave with $120^\circ$
order [Fig.~\ref{fig:pd}(d)]. Because the system is highly frustrated,
there is a strong possibility of intermediate phases. In
Fig.~\ref{fig:pd}(a), we show the phase diagram for 24-site clusters
as a function of $J_2 / J_1$, finding two intermediate phases: (1) a
quantum spin liquid and (2) an exotic spin wave state
[Fig.~\ref{fig:pd}(c)].

\begin{figure}[b]
  \centering
  \includegraphics*[height=\columnwidth,angle=-90,viewport=55 25 500 675,draft=false]{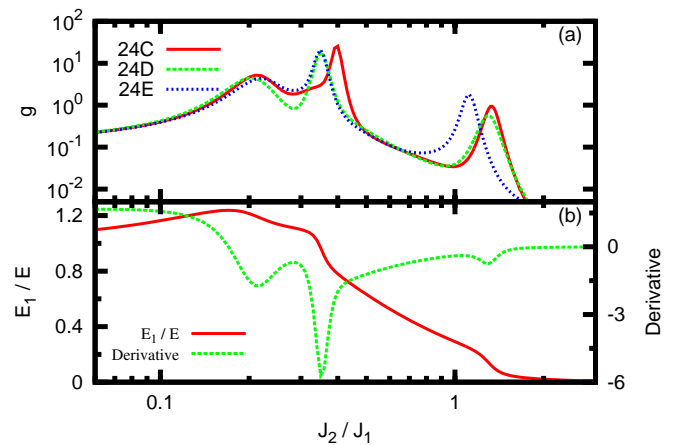}
  \caption{
    (a) Fidelity metric vs $J_2 / J_1$ for clusters $24C$, $24D$, and
    $24E$. (b) Ratio of the NN energy to the total energy $E_1 / E$ and its
    derivative (right axis) for the $24D$ cluster.
    \label{fig:fidelity}
  }
\end{figure}

To pin down the phase boundaries, we consider the ground-state
fidelity metric $g$, which has been shown to be an unbiased and
sensitive indicator of quantum phase
transitions~\cite{zanardi2006,rigol2009}. In
Fig.~\ref{fig:fidelity}(a), we show the fidelity metric for three
different 24-site clusters~\cite{varney2010,SupplMat}. There are three
peaks in $g$, indicating three quantum phase transitions. As we
discuss in greater detail below, the three clusters have slightly
different momentum space representations, resulting in the second and
third transitions to occur at slightly different values of $J_2 / J_1$
for each cluster.

Another indicator of a phase transition can be seen in the NN energy
$E_1$ (the NNN energy is denoted by $E_2$). In
Fig.~\ref{fig:fidelity}(b), we show the ratio $E_1 / E$ (here $E$ is
the ground-state energy) for the $24D$ cluster. The transition points
are directly connected with the inflection points in $E_1 / E$. To
demonstrate this more clearly, we also show the derivative of $E_1 /
E$, whose minima coincide with the transitions determined by the
fidelity metric.

\begin{figure}[t]
  \centering
  \includegraphics*[height=\columnwidth,angle=-90,viewport=20 0 350 725,draft=false]{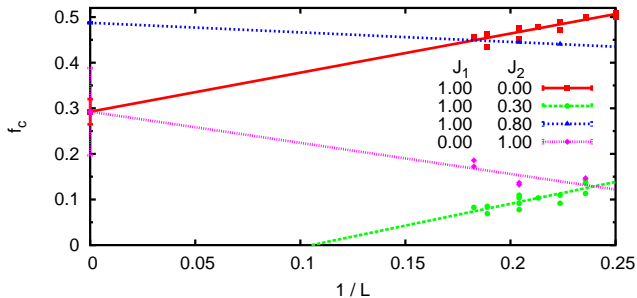}
  \caption{
    (Color online) Finite-size scaling of the condensate fraction $f_c$ for
    parameters that are representative of each phase depicted in
    Fig.~\ref{fig:pd} ($J_2 / J_1 = 0.00, 0.30, 0.80, \infty$). The color of
    each curve is consistent with the color coding in Fig.~\ref{fig:pd}. In
    the limit $L \to \infty$, the condensate fraction is nonzero in the
    antiferromagnetic, spin wave, and $120^\circ$ ordered phases.
    \label{fig:condensate}
  }
\end{figure}

From the mapping between spins and hard-core bosons, it follows that
the antiferromagnetic and the other two ordered states correspond to
Bose-Einstein condensates (BECs) in which bosons condense into quantum
states with different momenta. To characterize these phases, we
measure the condensate fraction $f_c = \Lambda_1 / N_b$ ($N_b$ is the
total number of bosons) by computing the largest eigenvalue
$\Lambda_1$ of the one-particle density matrix $\rho_{ij} = \braket{
  b_i^\dag b_j^{\phantom\dag}}$. If $f_c$ scales to a nonzero value in
the thermodynamic limit, then the system exhibits Bose-Einstein
condensation~\cite{penrose1956}. This is the case in three of our
phases, as depicted in Fig.~\ref{fig:condensate}.  For the Bose-metal
phase, on the other hand, the condensate fraction vanishes in the
thermodynamic limit, indicating the absence of BEC.

To further examine the properties of these phases, we calculate the
single-particle occupation at different momentum points
\begin{align}
  n({\bf k}) &= \braket{\alpha_{\bf k}^\dagger \alpha_{\bf
      k}^{\phantom\dag}} + \braket{\beta_{\bf k}^\dagger \beta_{\bf
      k}^{\phantom\dag}}.
\end{align}
Here, $\alpha_{\bf k}^{\phantom\dag}$ and $\beta_{\bf k}^{\phantom\dag}$ are
boson annihilation operators at momentum ${\bf k}$ for the $A$ and $B$
sublattices. Because we are studying finite-sized clusters, we utilize twisted
boundary conditions~\cite{poilblanc1991} and average over $40\times40$
boundary conditions to fully probe the Brillouin zone.

\begin{figure}[b]
  \centering
  \includegraphics[width=0.32\columnwidth,draft=false]{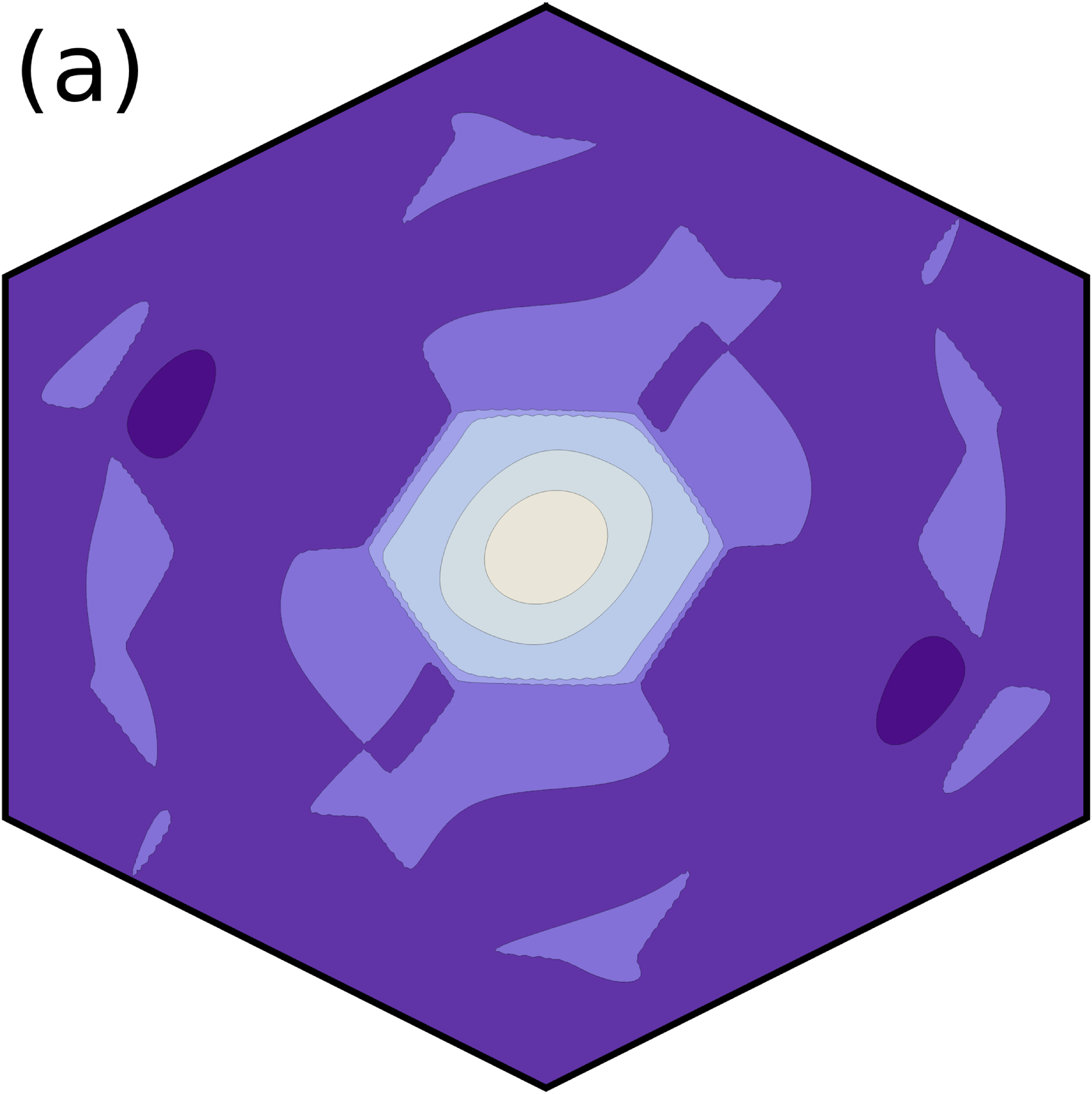}
  \includegraphics[width=0.32\columnwidth,draft=false]{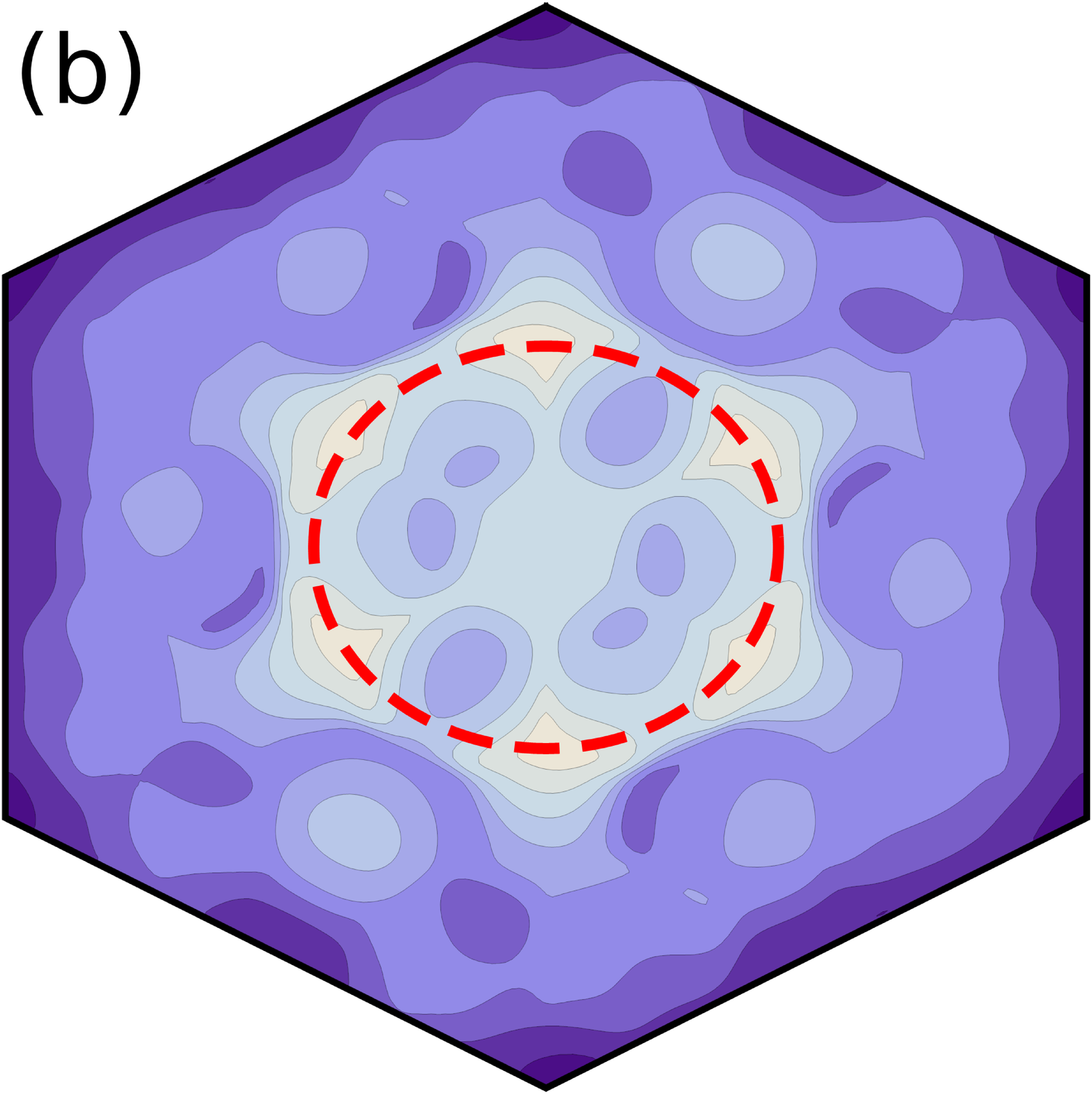}
  \includegraphics[width=0.32\columnwidth,draft=false]{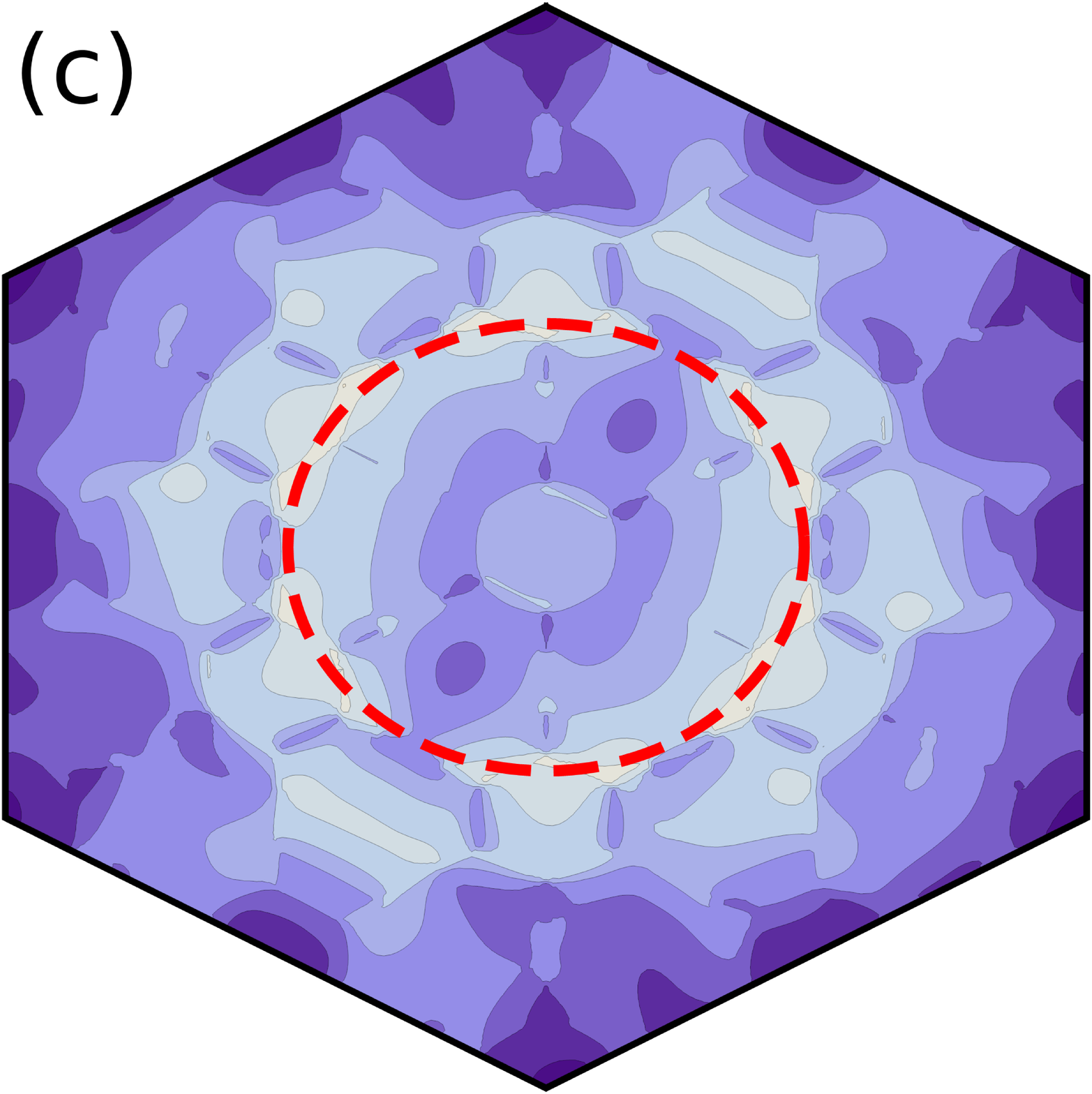}
  \includegraphics[width=0.32\columnwidth,draft=false]{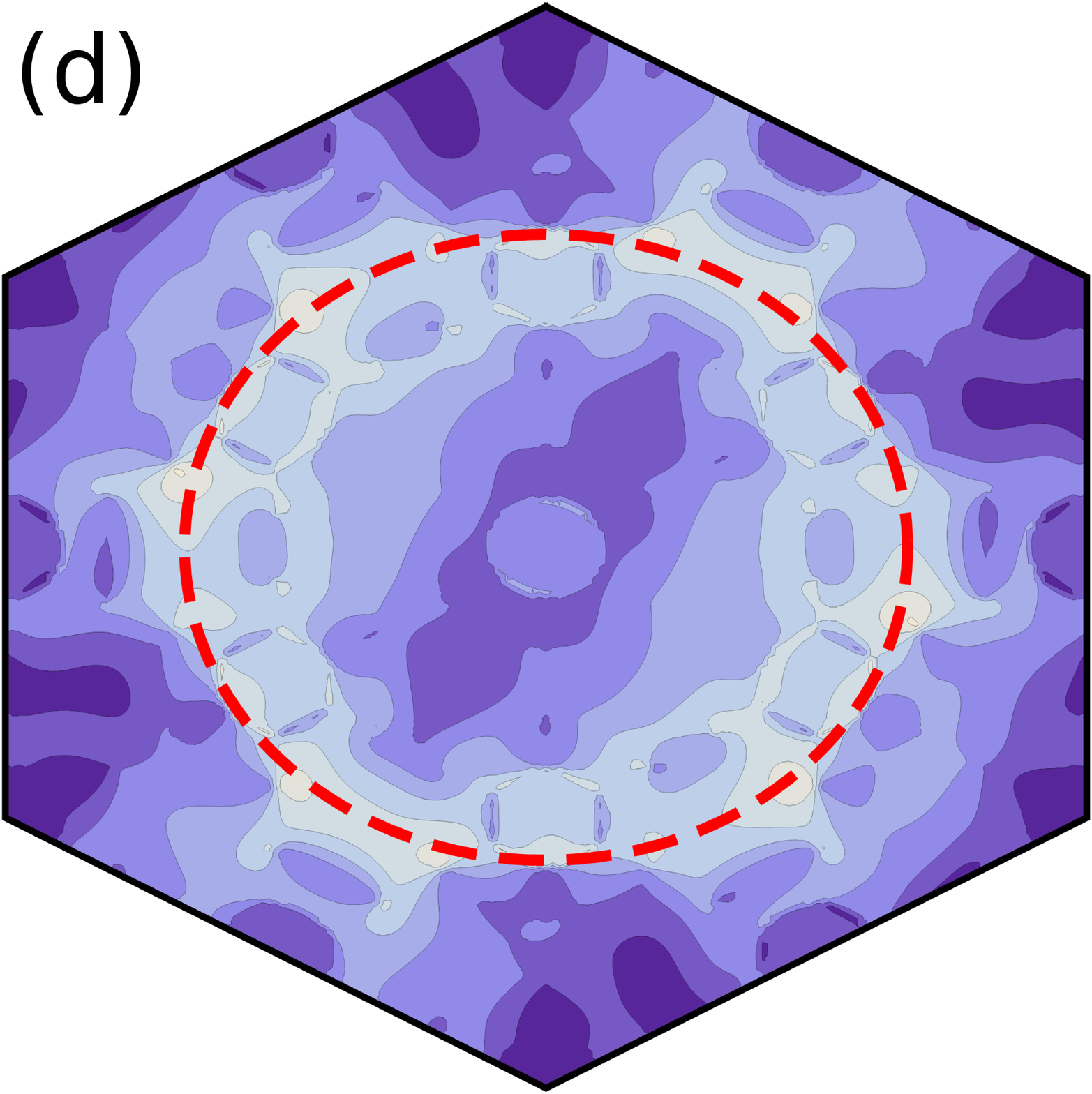}
  \includegraphics[width=0.32\columnwidth,draft=false]{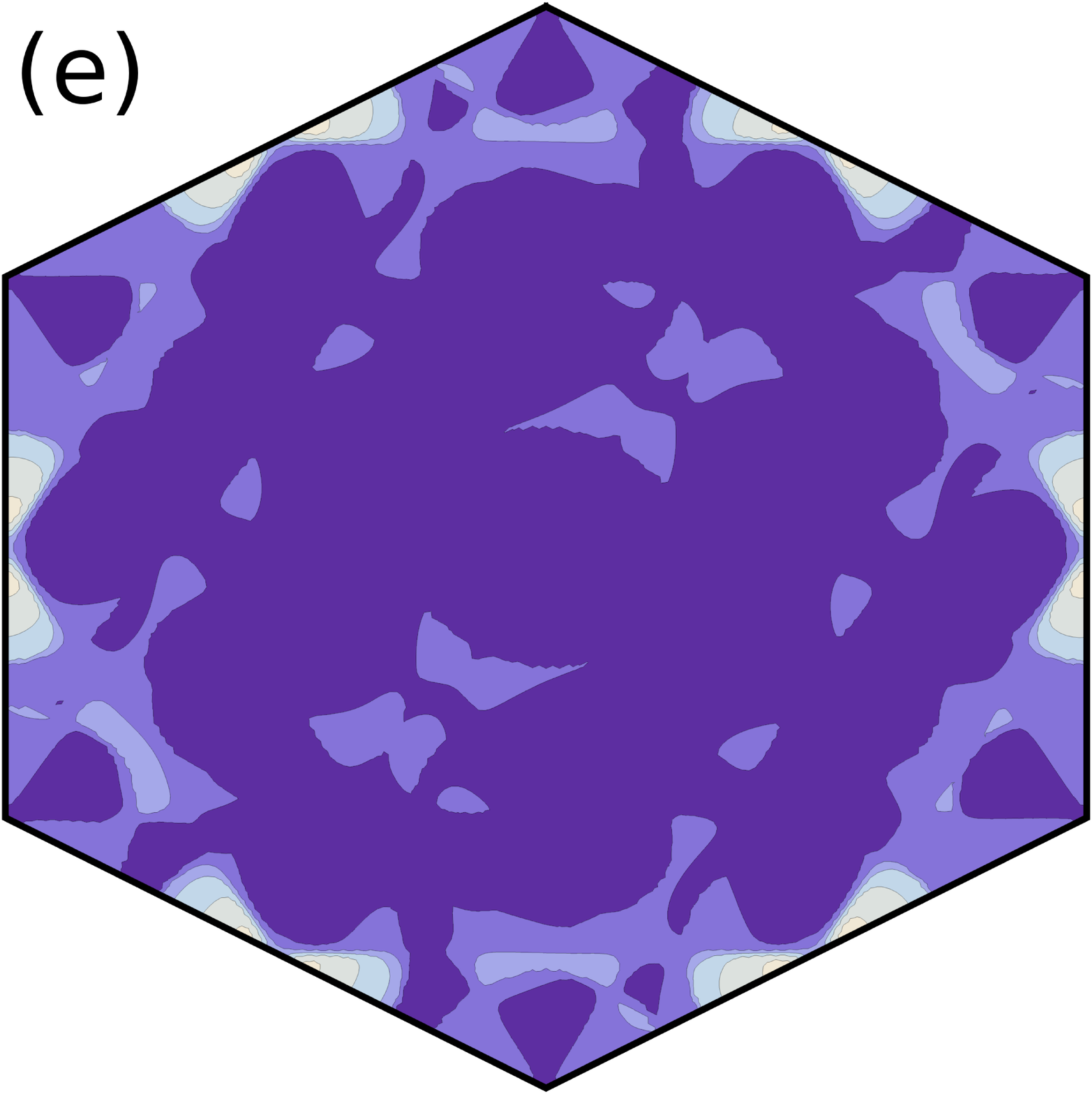}
  \includegraphics[width=0.32\columnwidth,draft=false]{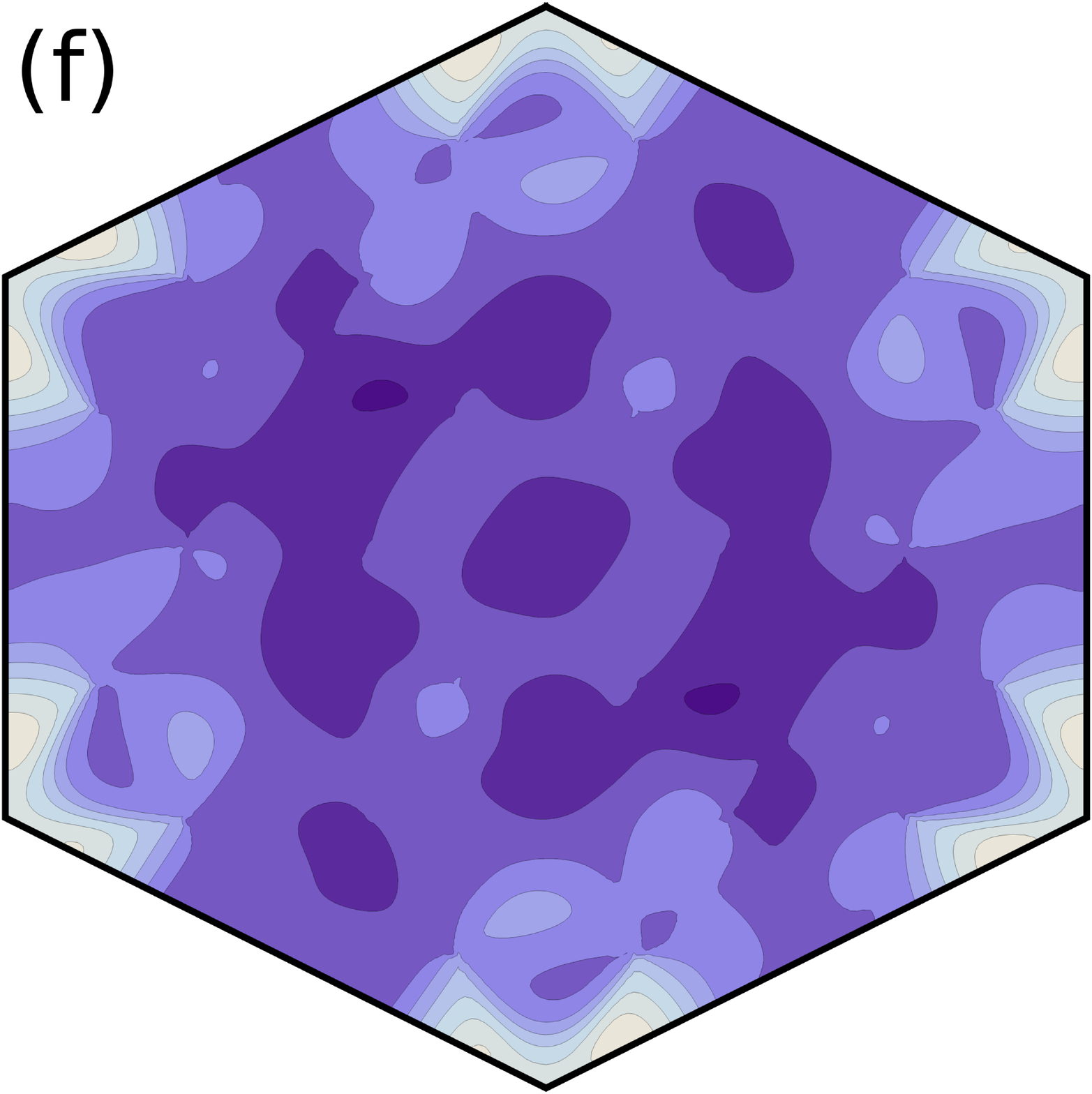}
  \caption{
    (Color online) $n({\bf k})$ vs ${\bf k}$ for (a) the N\'eel state
    ($J_2 / J_1 = 0$), (b-d) Bose metal ($J_2 / J_1 = 0.27, 0.30$,
    and $0.32$), (e) collinear spin wave ($J_2 / J_1 = 0.80$), and (f)
    the $120^\circ$ ordered state ($J_2 / J_1 = \infty$). In (b-d),
    the Bose surface is indicated by the dashed red line and 
    has  a radius of magnitude $k_B = 0.9$, $1.0$, and $1.4$,
    respectively. Each plot contains $19\,600$ $k$ points. 
   \label{fig:nk_contour}
 }
\end{figure}

In Fig.~\ref{fig:nk_contour}, we show the momentum distribution
function as a function of ${\bf k}$ for select values of $J_2 / J_1$
that are representative of each phase. In the first phase
[Fig.~\ref{fig:nk_contour}(a)], the momentum distribution function is
sharply peaked at ${\bf k} = \Gamma$, indicating an
antiferromagnetically ordered state. The third and fourth phases
[Figs.~\ref{fig:nk_contour}(e) and \ref{fig:nk_contour}(f),
respectively] also exhibit sharp peaks in $n({\bf k})$, but this time
at the edges of the Brillouin zone. For phase III
[Fig.~\ref{fig:nk_contour}(e)], $n({\bf k})$ is maximal at ${\bf k} =
M$, corresponding to the collinear spin wave state illustrated in
Fig.~\ref{fig:pd}(c). For phase IV [Fig.~\ref{fig:nk_contour}(f)],
$n({\bf k})$ is maximal at ${\bf k} = K$, as one would expect for a
$120^\circ$ ordered phase [Fig.~\ref{fig:pd}(d)].

The momentum distribution function in the Bose metal is depicted in
Figs.~\ref{fig:nk_contour}(b-d) for three different values of $J_2 /
J_1$. One can see there that $n({\bf k})$ in this phase exhibits a
remarkable $J_2 / J_1$-dependent Bose surface. Namely, the magnitude
of the Bose wave vector $k_B$ at which the maxima of $n({\bf k})$
occurs changes (increases) with increasing $J_2 / J_1$. The important
distinction to be made here is that those maxima do not reflect
Bose-Einstein condensation; i.e., they do not scale with the system
size as the ones in the other three phases do.

We should add that, in order to exclude other ordering tendencies in
phase II, we also examined the $S^z$ correlation function $C_{i,j} =
\braket{(S^z_{ia} - S^z_{ib})(S^z_{ja} - S^z_{jb})}$ and the
dimer-dimer correlation function $D_{ij,k\ell} = \braket{ ({\bf S}_i
  \cdot {\bf S}_j)({\bf S}_k \cdot {\bf S}_\ell)}$ and their
corresponding structure factors. Finite-size scaling of these
structure factors (not shown) made evident that neither charge density
wave formation nor dimer formation occurs. We also computed the
excitation gap in the Bose-metal phase and found it to be much smaller
than the (finite-size) excitation gap in the antiferromagnetic
state. In the antiferromagnetic phase, the system is gapless in the
thermodynamic limit, due to the spontaneous breaking of the
spin-rotation symmetry and the resulting Goldstone modes. Since the
gap in the Bose-metal phase is significantly smaller, we believe this
gap is also due to finite-size effects and will close in thermodynamic
limit. Our small system-sizes prevent us from reaching conclusive
results for this quantity after a finite-size
extrapolation. Nevertheless, all the phenomena we observed in this
phase are consistent with and indicate a Bose-metal phase.

\begin{figure}[bt]
  \centering
  \includegraphics*[height=\columnwidth,angle=-90,viewport=30 0 505 700,draft=false]{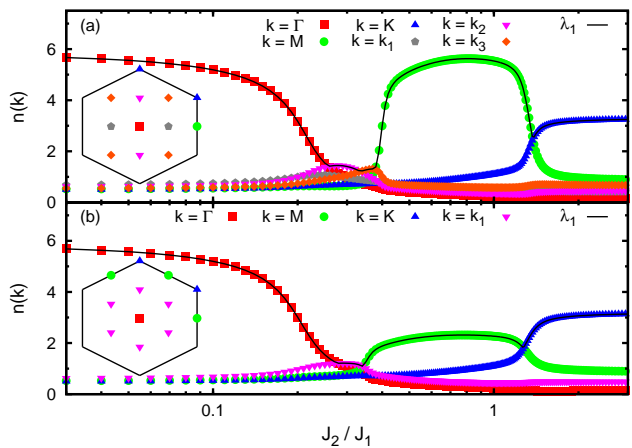}
  \caption{
    (Color online) Momentum distribution function (symbols) for the different
    values of ${\bf k}$ and clusters (a) $24C$ and (b) $24D$. Also
    shown are the largest eigenvalue $\lambda_1$ of the single-particle
    density matrix (line). The inset of each panel illustrates the k-points
    for each cluster. 
    \label{fig:nk_vs_t}
  }
\end{figure}

In Fig.~\ref{fig:nk_vs_t}, we illustrate how both the momentum
distribution function and the largest eigenvalue of
$\braket{\alpha_i^\dag \alpha_j^{\phantom\dag} + \beta_i^\dag
  \beta_j^{\phantom\dag}}$, $\lambda_1$, evolve over the entire
parameter space for two different 24-site clusters with periodic
boundary conditions. The maximum of $n({\bf k})$ perfectly matches
$\lambda_1$, and it is clear that the momenta of the condensates in
phases I, III, and IV are ${\bf k} = \Gamma$, $M$, and $K$,
respectively. In addition, it can be seen that the momentum
distribution in phase II exhibits a peak inside the Brillouin zone
that shifts to larger momenta as $J_2 / J_1$ is increased.

Phases III and IV exhibit an interesting phenomenon that can be
unveiled by examining the degeneracy of the largest eigenvalues of the
one-particle density matrix. In an ordinary BEC state, condensation
occurs to a unique effective single-particle state, and thus the
largest eigenvalue of the density matrix is nondegenerate and
$O(N_b)$, while the second largest eigenvalue is already $O(1)$. This
is what we find in the antiferromagnetic state (phase I). However, in
general, condensation can occur to more than one effective
one-particle state~\cite{leggett2001,stanescu2008}, and various
largest eigenvalues of the one-particle density matrix may become
$O(N_b)$ and degenerate. This fragmentation occurs in phases III and
IV. For phase IV, it is trivial to realize that the condensate must be
degenerate in the limit $J_2/J_1\rightarrow\infty$, where the system
consists of two disconnected triangular lattices. Interestingly, in
this model, fragmentation occurs for all values of $J_2/J_1$ in phases
III and IV, and is related to the number of $M$ or $K$ points present
in the clusters under consideration.

In summary, we have studied a frustrated $XY$ model on a honeycomb
lattice. We find that this model exhibits four phases [see phase
diagram in Fig.~\ref{fig:pd}(a)]: (I) a BEC at ${\bf k} = \Gamma$
(antiferromagnetism), (II) a Bose metal (spin liquid), (III) a BEC at
${\bf k} = M$ (a collinear spin wave), and (IV) a BEC at ${\bf k} = K$
($120^\circ$ order). The Bose-metal phase is characterized by a
parameter dependent peak in $n({\bf k})$ and a lack of condensation,
solid order, and dimer order. This work provides the first convincing
example of a gapless spin liquid in a surprisingly simple model of
$XY$ spins. We believe that there is no fundamental challenge to
realize the Bose-metal phase in experimental systems dealing with
spins or ultracold optically trapped bosons in the regime of large
on-site Hubbard repulsion. Finally, recent experimental~\cite{yan2011}
and theoretical~\cite{meng2010,mulder2010,clark2010,albuquerque2011}
work suggest that exotic quantum spin liquids might exist in related
lattice models for $SU(2)$ spins. It would be interesting to see what
relationship, if any, exists between these possible $SU(2)$ spin
liquids and our Bose metal, which can be studied by introducing
interaction terms for the bosons.

This research was supported by NSF through JQI-PFC (C.N.V. and K.S.),
ONR (C.N.V. and M.R.), U.S.-ARO (V.G.), and NSF under Grant No.\
PHY05-51164. The authors thank M.P.A. Fisher and L. Fu for
discussions, A. L\"auchli for bringing to our attention the collinear
nature of phase III, and M.M. Ma\'ska for useful suggestions on the
manuscript.

\bibliography{spin_liquid9}

\begin{thebibliography}{34}
\bibitem [{not()}]{note1}%
  \BibitemOpen
  \href@noop {} {}\bibinfo {note} {Note that since spin-$1/2$ models can be
  mapped onto hard-core boson models and vice versa, one can use the terms
  ``spin liquid'' and ``Bose liquid'' (also ``Bose metal'') interchangeably
  when referring to such states.}\BibitemShut {Stop}%
\end{thebibliography}

\end{document}